\documentstyle[twocolumn,twoside,floats,prb,aps,psfig]{revtex}

\begin{document}
\newcommand{\beq}{\begin{equation}}
\newcommand{\eeq}{\end{equation}}
\newcommand{\degree}{$^{\rm\circ}$}

\title{A-Tract Induced DNA Bending is a Local Non-Electrostatic Effect}

\author{Alexey K. Mazur}

\address{Laboratoire de Biochimie Th\'eorique, CNRS UPR9080\\
Institut de Biologie Physico-Chimique\\
13, rue Pierre et Marie Curie, Paris,75005, France.\\
FAX: +33[0]1.58.41.50.26 Email: alexey@ibpc.fr}

\date{\today}
\maketitle

\begin{abstract}

The macroscopic curvature induced in double helical B-DNA by regularly
repeated adenine tracts (A-tracts) is a long known, but still
unexplained phenomenon. This effect plays a key role in DNA studies
because it is unique in the amount and the variety of the available
experimental information and, therefore, is likely to serve as a gate
to the unknown general mechanisms of recognition and regulation of
genome sequences. The dominating idea in the recent years was that, in
general, macroscopic bends in DNA are caused by long range
electrostatic repulsion between phosphate groups when some of them are
neutralized by proximal external charges. In the case of A-tracts this
may be specifically bound solvent counterions. Here we report about
molecular dynamics simulations where a correct static curvature in a
DNA fragment with phased adenine tracts emerges spontaneously in
conditions where any role of counterions or long range electrostatic
effects can be excluded.

\end{abstract}

\section*{Results and Discussion}

Although the macroscopic curvature of DNA induced by adenine-tracts
(A-tracts) was discovered almost two decades ago
\cite{Marini:82,Wu:84} structural basis for this phenomenon remains
unclear. A few models considered originally
\cite{Trifonov:80,Levene:83,Calladine:88}
suggested that it is caused by intrinsic conformational preferences of
certain sequences, but all these and similar theories 
failed to explain experimental data obtained later. \cite{Crothers:99}
Calculations show that the B-DNA duplex is mechanically
anisotropic, \cite{Zhurkin:79} that bending towards minor grooves
of some A-tracts is strongly facilitated, \cite{Sanghani:96}
and that the macroscopic curvature becomes
energetically preferable once the characteristic
A-tract structure is maintained by freezing or imposing constraints.
\cite{Kitzing:87,Chuprina:88,Zhurkin:91} However, the static curvature
never appears spontaneously in calculations unbiased {\em a priori}
and these results leave all doors open for the possible physical
origin of the effect. In the recent years the main attention has been
shifted to specific interactions between DNA and solvent counterions
that can bend the double helix by specifically neutralizing some
phosphate groups. \cite{Mirzabekov:79,Levene:86,Strauss:94,Travers:95,McFail-Isom:99}
The possibility of such mechanism is often evident in protein-DNA
complexes, and it has also been demonstrated by direct chemical
modification of a duplex DNA. \cite{Strauss:94} In the case of the free
DNA in solution, however, the available experimental observations are
controversial. \cite{McFail-Isom:99,Chiu:99} Molecular dynamics
simulations of a B-DNA in an explicit counterion shell could neither
confirm nor disprove this hypothesis. \cite{Young:98}
Here we report the first example where stable static curvature
emerges spontaneously in molecular dynamics simulations. Its direction
is in striking agreement with expectations based upon experimental
data. However, we use a minimal B-DNA model without counterions,
which strongly suggests that they hardly play a key role
in this effect.

Figure \ref{FTj1} exhibits results of a 10 ns simulation of dynamics
of a 25-mer B-DNA fragment including three A-tracts separated by one
helical turn. This sequence has been constructed after many
preliminary tests with shorter sequence motives. Our general strategy
came out from the following considerations. Although the A-tract
sequences that induce the strongest bends are known from experiments,
probably not all of them would work in simulations. There are natural
limitations, such as the precision of the model, and, in addition,
the limited duration of trajectories may be insufficient
for some A-tracts to adopt their specific conformation. Also, we can
study only short DNA fragments, therefore, it is preferable to place A-tracts
at both ends in order to maximize the possible bend. There is,
however, little experimental evidence of static curvature in short DNA
fragments, and one may well expect the specific A-tract structure
to be unstable near the ends. That is why we did
not simply take the strongest experimental ``benders'', but looked for
sequence motives that in calculations readily adopt the characteristic
local structure, with a narrow minor groove profile and high propeller
twist, both in the middle and near the ends of the duplex. The
complementary duplex $\rm AAAATAGGCTATTTTAGGCTATTTT$ has been constructed by
repeating and inverting one such motive.

\begin{figure}
\centerline{\psfig{figure= 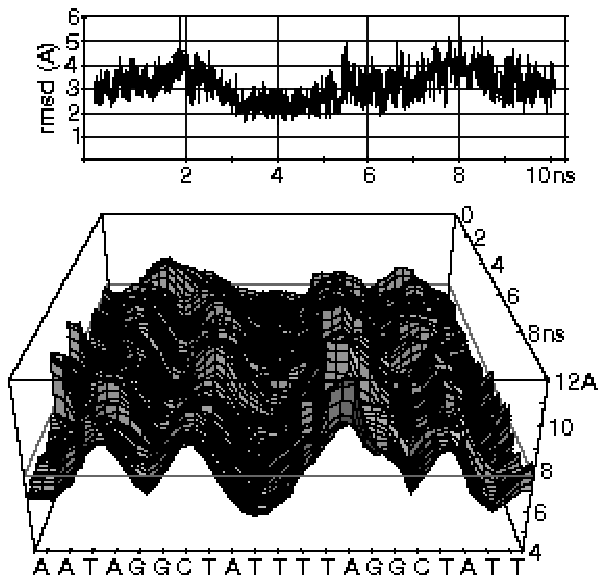,height=8cm,angle=0.,%
bbllx=120bp,bblly=120bp,bburx=320bp,bbury=310bp,clip=t}}
\caption{(a)}
\end{figure}\addtocounter{figure}{-1}

\begin{figure}
\centerline{\psfig{figure= 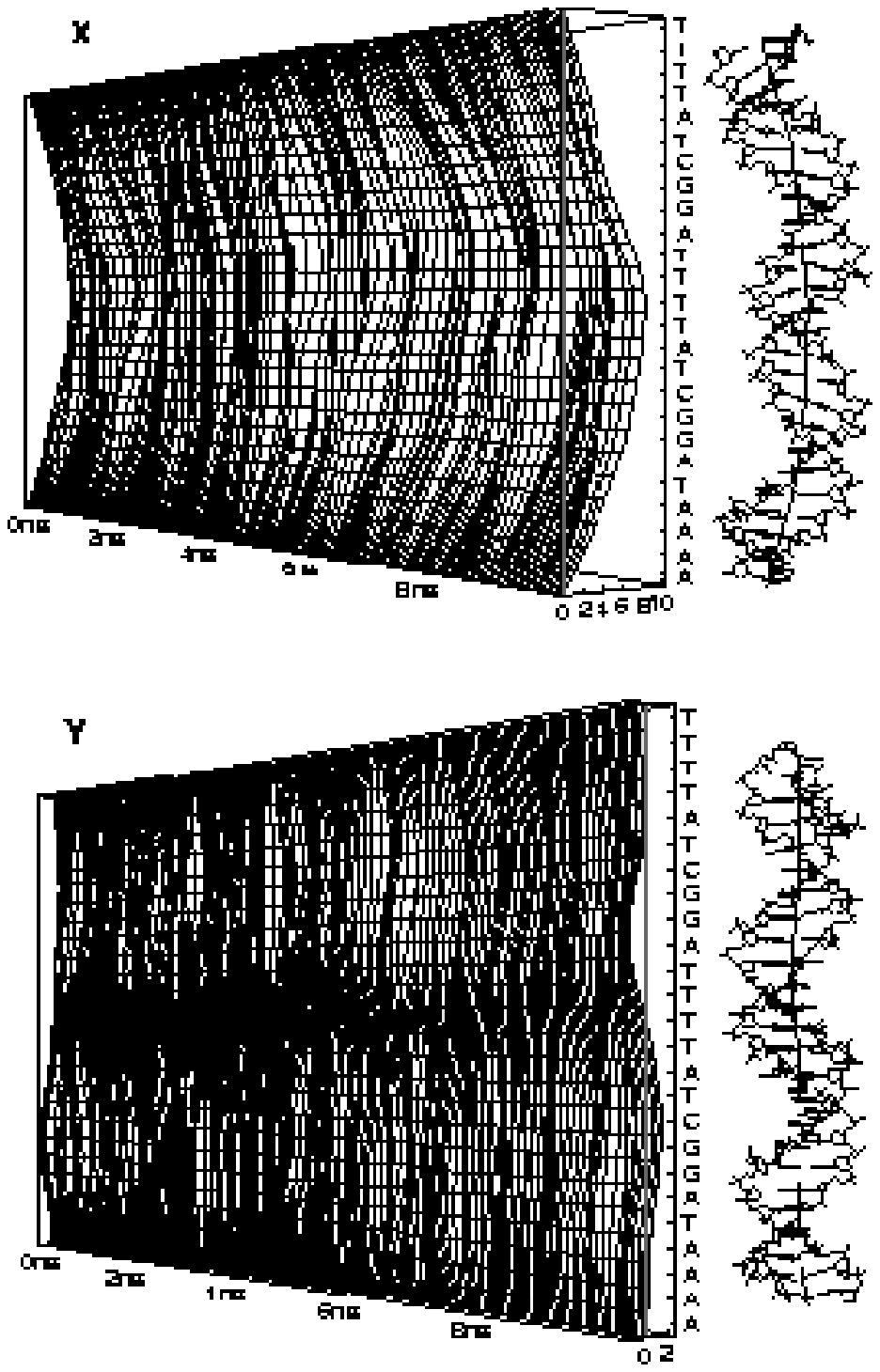,height=12cm,angle=0.,%
bbllx=0bp,bblly=0bp,bburx=280bp,bbury=450bp,clip=t}}
\caption{(c)}
\end{figure}\addtocounter{figure}{-1}

\begin{figure}
\centerline{\psfig{figure= 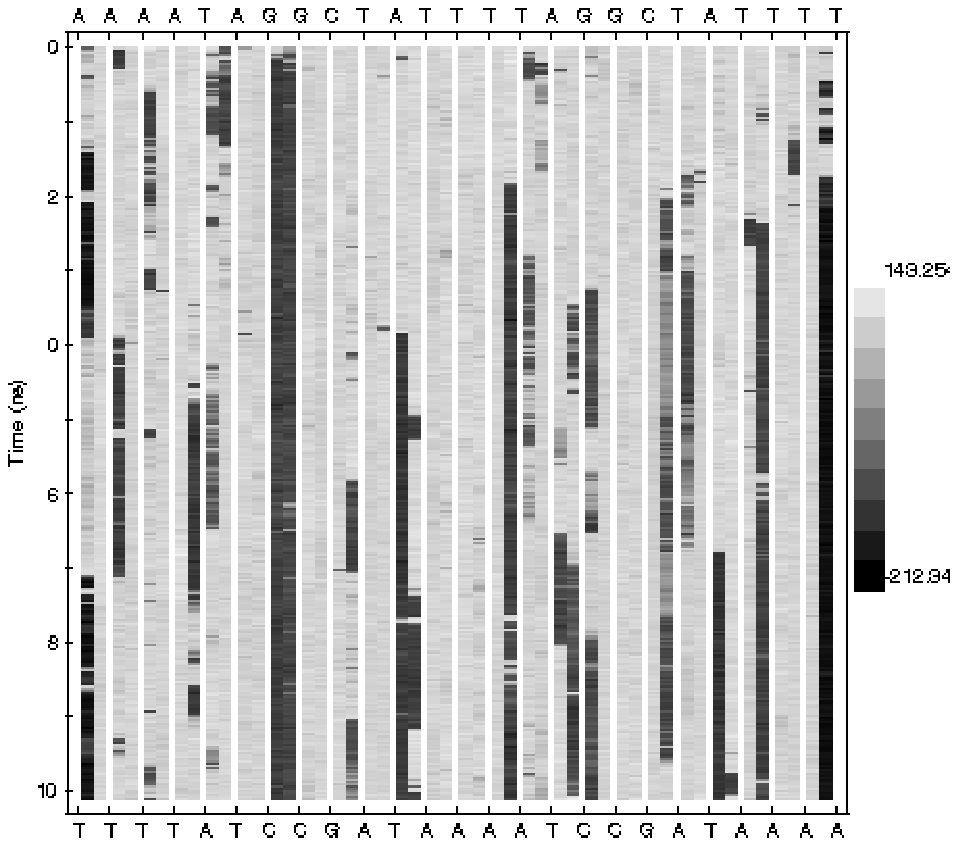,height=8cm,angle=0.,%
bbllx=50bp,bblly=90bp,bburx=370bp,bbury=350bp,clip=t}}
\caption{(b)}
\end{figure}\addtocounter{figure}{-1}

\begin{figure}
\centerline{\psfig{figure=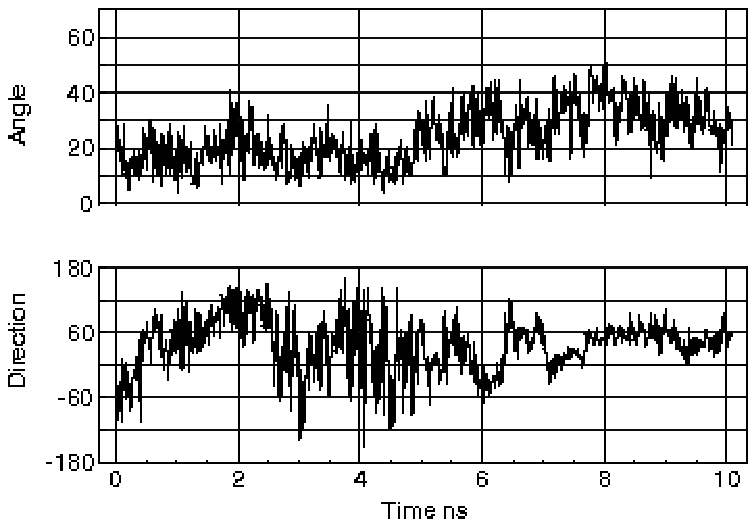,height=6cm,angle=0.,%
bbllx=50bp,bblly=80bp,bburx=340bp,bbury=250bp,clip=t}}
\caption{\label{FTj1}
Representative results from the first 10 ns MD simulation of a 25-mer
double helical DNA fragment. (a) The time variation of the heavy atom
rmsd from the canonical B-DNA form and the evolution of the profile of
the minor groove. (b) Dynamics of  $\rm B_I\leftrightarrow B_{II}$
backbone transitions. (c) The time evolution of the optimal helical
axis, which is a best fit axis of coaxial cylindrical surfaces passing
through sugar atoms. In all cases considered here it is very close to
that produced by the Curves algorithm. \protect\cite{Curves:} Two
perpendicular projections are shown with the  corresponding views of
the average conformation during the last nanosecond shown on the
right.  (d) The time variation the bending angle and direction. The
bending angle is measured between the two ends of the optimal helical
axis. The bending direction is characterized by the angle between the
X-projection plane in plate (c) and the $xz$ plane of the local DNA
coordinate frame constructed in the center of the duplex according to
the Cambridge convention \protect\cite{Dickerson:89}.
}
\end{figure}

The upper trace in plate (a) shows the time dependence
of rmsd from the canonical B-DNA model. It fluctuates below 4 \AA\
sometimes falling down to 2 \AA, which is very low for the double
helix of this length indicating that all helical parameters are well
within the range of the B-DNA family. The lower surface plot shows
the time evolution of the minor DNA groove. The surface is formed by
75 ps time-averaged successive minor groove profiles, with that on
the front face corresponding to the final DNA conformation. The groove
width is evaluated by using space traces of C5' atoms as described
elsewhere \cite{Mzjmb:99}. Its value is given in angstr\"oms and the
corresponding canonical B-DNA level of 7.7 \AA\ is marked by the
straight dotted lines on the faces of the box. It is seen that the
overall groove shape has established after 2 ns and
remained stable later, with noticeable local fluctuations. In all
A-tracts the groove strongly narrows towards 3' ends and widens
significantly at the boundaries. There are two less significant
relative narrowings inside non A-tract sequences as well.

Dynamics of $\rm B_I\leftrightarrow B_{II}$ backbone transitions are
shown in plate (b). The B$_{\rm I}$ and B$_{\rm II}$ conformations are
distinguished by the values of two consecutive backbone torsions,
$\varepsilon$ and $\zeta$. In a transition they change concertedly
from (t,g$^-$) to  (g$^-$,t). The difference $\zeta -\varepsilon$ is,
therefore, positive in B$_{\rm I}$ state and negative in B$_{\rm II}$,
and it is used in Fig. (d) as a monitoring indicator, with the
corresponding gray scale levels shown on the right. Each base pair
step is characterized by a column consisting of two sub-columns, with
the left sub-columns referring to the sequence written at the top in
5'-3' direction from left to right. The right sub-columns refer to the
complementary sequence shown at the bottom. It is seen that, in
A-tracts, the B$_{\rm II}$ conformation is preferably found in ApA
steps and that $\rm B_I\leftrightarrow B_{II}$ transitions in
neighboring steps often occur concertedly so that along a single
A-strand $\rm B_I$ and $\rm B_{II}$ conformations tend to alternate.
The pattern of these transitions reveals rather slow dynamics and
suggests that MD trajectories in the 10 ns time scale are still not
long enough to sample all relevant conformations. Note, for instance,
a very stable $\rm B_{II}$ conformation in both strands at one of the
GpG steps.

Plate (c) shows the time evolution of the overall shape of the helical
axis. The optimal curved axes of all DNA conformations saved during
dynamics were rotated with the two ends fixed at the OZ axis to put
the middle point at the OX axis. The axis is next characterized by two
perpendicular projections labeled X and Y. Any time section of the
surfaces shown in the figure gives the corresponding axis projection
averaged over a time window of 75 ps. The horizontal deviation is
given in angstr\"oms and, for clarity, its relative scale is two times
increased with respect to the true DNA length. Shown on the right are
two perpendicular views of the last one-nanosecond-average
conformation. Its orientation is chosen to correspond approximately that
of the helical axis in the surface plots.

It is seen that the molecule maintained a planar bent shape during a
substantial part of the trajectory, and that at the end the bending
plane was passing through the three A-tracts. The X-surface clearly
shows an increase in bending during the second half of the trajectory.
In the perpendicular Y-projection the helical axis is locally wound,
but straight on average. The fluctuating pattern in Y-projection
sometimes reveals two local maxima between A-tracts, which corresponds
to two independent bends with slightly divergent directions. One may
note also that there were at least two relatively long periods when
the axis was almost straight, namely, around 3 ns and during the fifth
nanosecond. At the same time, straightening of only one of the two
bending points is a more frequent event observed several times in the
surface plots.

Finally, plate (d) shows the time fluctuations of the bending
direction and angle. The bending direction is characterized by the
angle between the X-projection plane in plate (c) and the $xz$ plane
of the local DNA coordinate frame constructed in the center of the
duplex. According to the Cambridge convention \cite{Dickerson:89} the
local $x$ direction points to the major DNA groove along the short axis
of the base-pair, while the local $z$ axis direction is adjacent to
the optimal helicoidal axis. Thus, a zero angle between
the two planes corresponds to the overall bend to the
minor groove exactly at the central base pair. In both plots, short
time scale fluctuations are smoothed by averaging with a window of 15
ps. The total angle measured between the opposite axis ends
fluctuates around 10-15\degree\ in the least bent states and raises to
average 40-50\degree\ during periods of strong bending. The maximal
instantaneous bend of 58\degree\ was observed at around 8 ns.

The bending direction was much more stable during the last few
nanoseconds, however, it fluctuated at a roughly constant value of
50\degree starting from the second nanosecond. This value means that
the center of the observed planar bend is shifted by approximately two
steps from the middle base pair so that its preferred direction is to
the minor groove at the two ATT triplets, which is well distinguished
in plate (c) as well, and corresponds to the local minima in the minor
groove profiles in plate (a). During the periods when the molecule
straightened the bending direction strongly fluctuates. This effect is
due to the fact that when the axis becomes straight the bending plane
is not defined, which in our case appears when the central point of
the curved axis passes close to the line between its ends. It is very
interesting, however, that after the straightening, the bending is
resumed in approximately the same direction.

Figure \ref{FTj2} exhibits similar data for another 10 ns trajectory
of the same DNA fragment, computed in order to check reproducibility
of the results. A straight DNA conformation was taken from the initial
phase of the previous trajectory, energy minimized, and restarted with
random initial velocities. It shows surprisingly similar results as
regards the bending direction and dynamics in spite of a somewhat
different minor groove profile and significantly different
distribution of $\rm B_I$ and $\rm B_{II}$ conformers along the
backbone. Note that in this case the helical axis was initially
S-shaped in X-projection, with one of the A-tracts exhibiting a
completely opposite bending direction. Fluctuations of the bending
direction are reduced and are similar to the final part of the first
trajectory, which apparently results from the additional re-equilibration.
In this case the maximal instantaneous bend of 71\degree\ was
observed at around 4 ns.

\begin{figure}
\centerline{\psfig{figure= 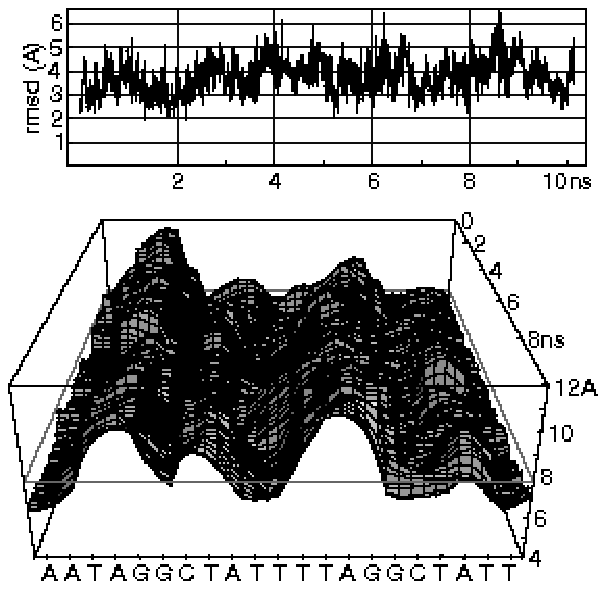,height=8cm,angle=0.,%
bbllx=120bp,bblly=120bp,bburx=320bp,bbury=320bp,clip=t}}
\caption{(a)}
\end{figure}\addtocounter{figure}{-1}

\begin{figure}
\centerline{\psfig{figure= 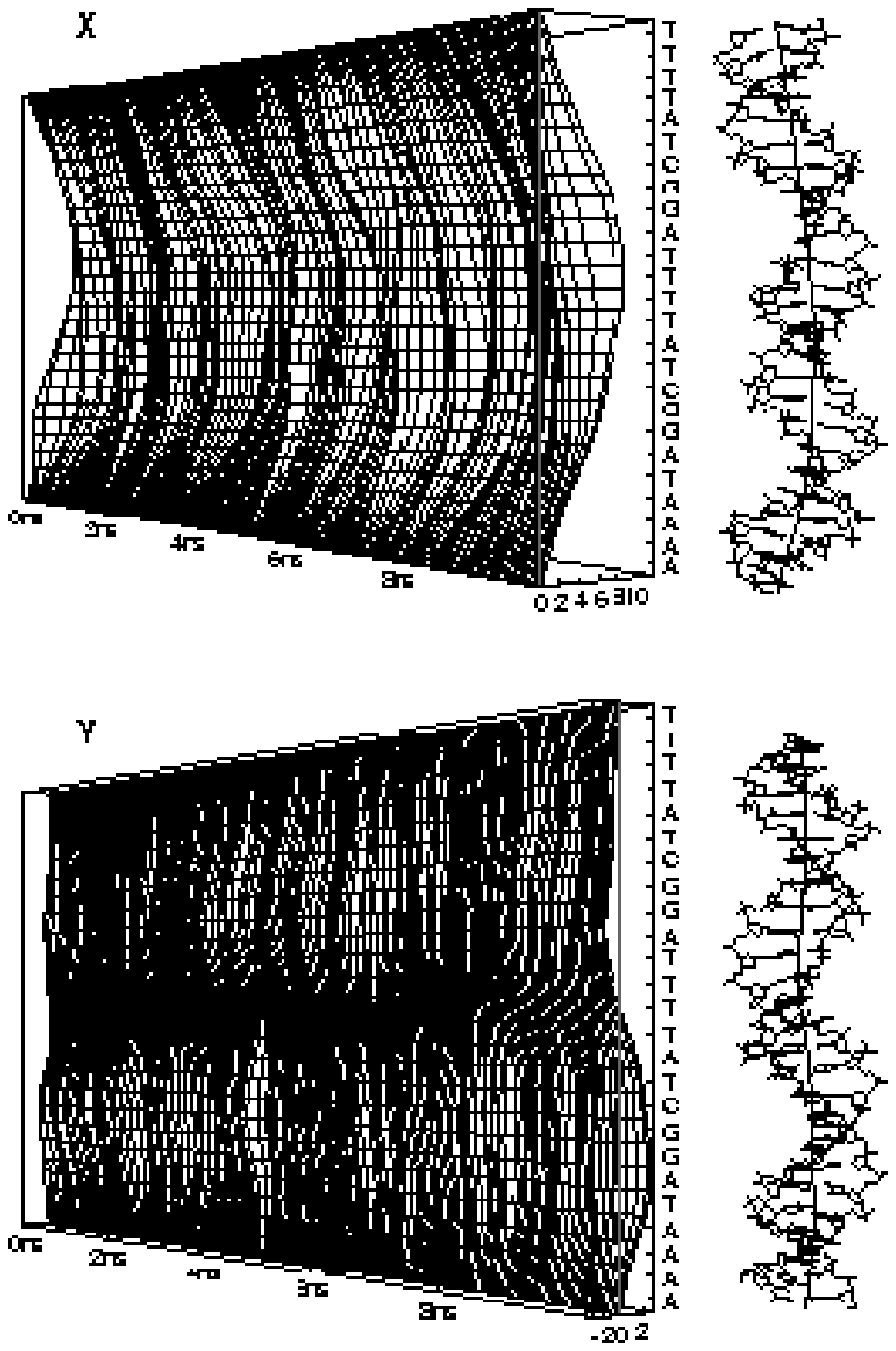,height=12cm,angle=0.,%
bbllx=0bp,bblly=0bp,bburx=280bp,bbury=450bp,clip=t}}
\caption{(c)}
\end{figure}\addtocounter{figure}{-1}

\begin{figure}
\centerline{\psfig{figure= 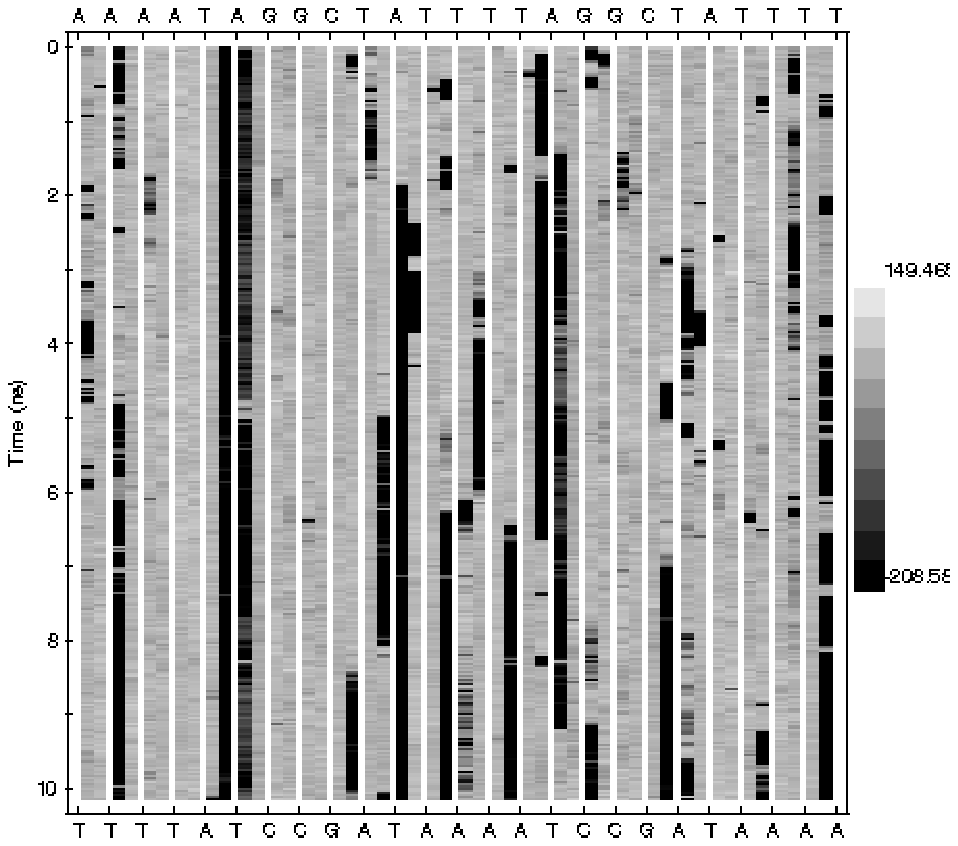,height=8cm,angle=0.,%
bbllx=50bp,bblly=90bp,bburx=370bp,bbury=350bp,clip=t}}
\caption{(b)}
\end{figure}\addtocounter{figure}{-1}

\begin{figure}
\centerline{\psfig{figure=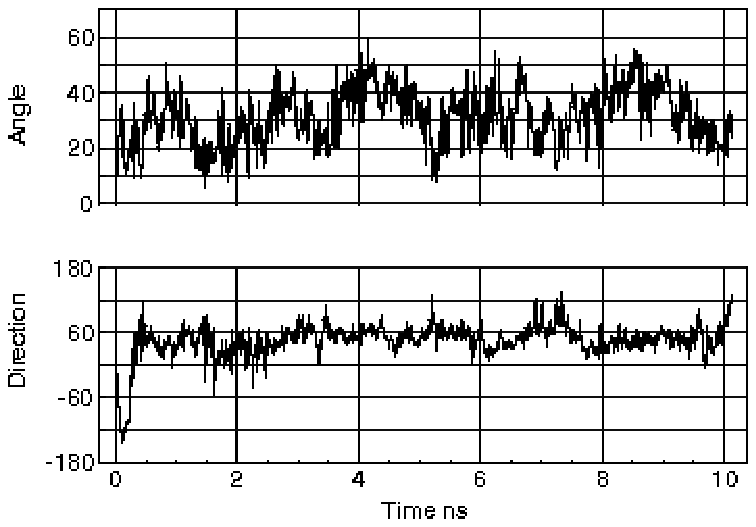,height=6cm,angle=0.,%
bbllx=50bp,bblly=80bp,bburx=340bp,bbury=260bp,clip=t}}
\caption{\label{FTj2}
Representative results from the second 10 ns MD trajectory of the same
DNA fragment. Notation as in Fig. \protect\ref{FTj1}.
}
\end{figure}

\begin{figure}
\centerline{\psfig{figure=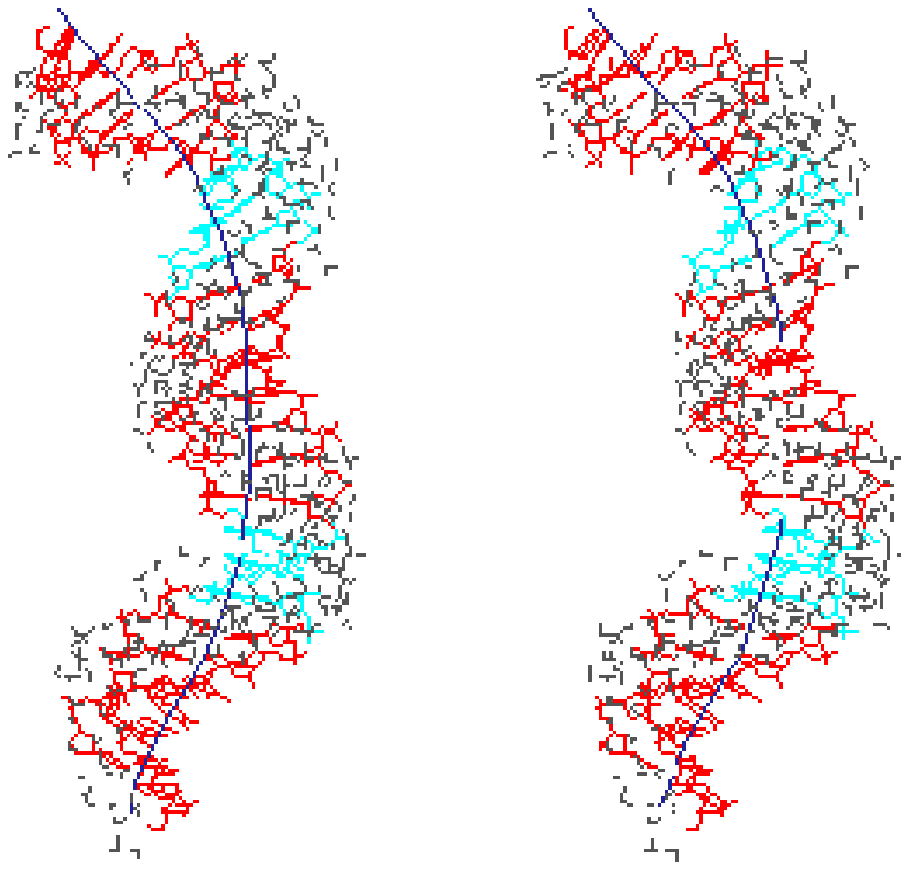,height=10cm,angle=0.,%
bbllx=160bp,bblly=160bp,bburx=440bp,bbury=440bp,clip=t}}
\caption{\label{FSnap}
A stereo snapshot of the system at around 8.5 ns of the second
trajectory. AT base pairs are shown in red and GC base pairs in blue.
}
\end{figure}

Comparison of traces in plates (a) and (d) in Figs. \ref{FTj1} and
\ref{FTj2} clearly shows that large scale slow fluctuations of rmsd
are caused by bending. The rmsd drops down to 2 \AA\ when the duplex
is straight and raises beyond 6 \AA\ in strongly bent conformations.
In both trajectories the molecule experienced many temporary transitions to
straight conformations which usually are very short living. These
observations suggest that the bent state is relatively more stable
than the straight one and, therefore, the observed behavior
corresponds to static curvature. In conformations averaged over
successive one nanosecond intervals the overall bending angle is
35-45\degree\ except for a few periods in the first trajectory. Figure
\ref{FSnap} shows a snapshot from around 8.5 ns of the second
trajectory where the rmsd from the straight canonical B-DNA reached
its maximum of 6.5 \AA. The strong smooth bent towards the minor
grooves of the three A-tracts is evident, with the overall bending
angle around 61\degree .

All transformations exhibited in Figs. \ref{FTj1} and \ref{FTj2} are
isoenergetic, with the total energy fluctuating around the same level
established during the first nanosecond already, and the same is true
for the average helicoidal parameters. Plates (b), however, indicate
that there are much slower motions in the system, and this observation
precludes any conclusions concerning the global stability of the
observed conformations. Moreover, we have computed yet another
trajectory for the same molecule starting from the canonical A-DNA
form. During 10 ns it converged to a similarly good B-DNA structure
with the same average total energy, but the bending pattern was not
reproduced. It appears, therefore, that the conformational space is
divided into distinct domains, with transitions between them probably
occurring in much longer time scales. However, the very fact that the
stable curvature in good agreement with experimental data emerges in
trajectories starting from a featureless straight canonical B-DNA
conformation strongly suggests that the true molecular mechanism of
the A-tract induced bending is reproduced. Therefore, it cannot depend
upon the components discarded in our calculations, notably, specific
interactions with solvent counterions and long-range electrostatic
effects.

We are not yet ready to present a detailed molecular mechanism
responsible for the observed curvature because even in this relatively
small system it is difficult to distinguish the cause and the
consequences. We believe, however, that all sorts of bending of the
double helical DNA, including that produced by ligands and that due to
intrinsic sequence effects, have its limited, but high flexibility as a common
origin. Its own conformational energy has the global minimum in a
straight form, but this minimum is very broad and flat, and DNA
responds by distinguishable bending to even small perturbations. The
results reported here prove that in the case of A-tracts these
perturbations are produced by DNA-water interactions in the minor
groove. Neither long range phosphate repulsion nor counterions are
essential. The curvature is certainly connected with the specific
A-tract structure and modulations of the minor groove width, but it
does not seem to be strictly bound to them. In dynamics,
conformations, both smoothly bent and kinked at the two insertions between
the A-tracts, are observed periodically. Note also, that the minor
groove profile somewhat differs between the two trajectories and that
it does not change when the molecule straightens. We strongly believe, however,
the experimental data already available will finally allow one to solve this
problem by theoretical means, including the approach described here, and
we continue these attempts.

\section*{Methods}

Molecular dynamics simulations have been performed with the internal
coordinate method (ICMD) \cite{Mzjbsd:89,Mzjcc:97} including special
technique for flexible sugar rings \cite{Mzjchp:99}. The so-called
``minimal B-DNA'' model was used \cite{Mzjacs:98,Mzlanl:99} which
consists of a double helix with the minor groove filled with explicit
water. Unlike the more widely used models, it does not involve
explicit counterions and damps long range electrostatic interactions
in a semi-empirical way by using distance scaling of the electrostatic
constant and reduction of phosphate charges. The DNA model was same
as in earlier reports, \cite{Mzjacs:98,Mzlanl:99} namely, all torsions
were free as well as bond angles centered at sugar atoms, while other
bonds and angles were fixed, and the bases held rigid. AMBER94
\cite{AMBER94:,Cheatham:99} force field and atom parameters were used
with TIP3P water \cite{TIP3P:} and no cut off schemes.
With a time step of 10 fs, these simulation conditions require around
75 hours of cpu per nanosecond on a Pentium II-200 microprocessor.

The initial conformations were prepared by vacuum energy minimization
starting from the fiber B-DNA model constructed from the published atom
coordinates. \cite{Arnott:72} The subsequent hydration protocol to
fill up the minor groove \cite{Mzjacs:98} normally adds around 16
water molecules per base pair. The heating and equilibration protocols
were same as before \cite{Mzjacs:98,Mzlanl:99}. During the runs, after
every 200 ps, water positions were checked in order to identify those
penetrating into the major groove and those completely separated.
These molecules, if found, were removed and next re-introduced in
simulations by putting them with zero velocities at random positions
around the hydrated duplex, so that they could readily re-join
the core system. This procedure assures stable conditions, notably, a
constant number of molecules in the minor groove hydration cloud and
the absence of water in the major groove, which is necessary for fast
sampling \cite{Mzlanl:99}. The interval of 200 ps between the checks
is small enough to assure that on average less then one molecule is
repositioned and, therefore, the perturbation introduced is considered
negligible.

\section*{Acknowledgements} {I thank R. Lavery for useful discussions
as well as critical comments and suggestions to the paper. }

\section*{Appendix}
This section contains comments from anonymous referees of peer-review
journals where the manuscript has been considered for publication, but
rejected.
\subsection{Journal of Molecular Biology}

\subsubsection {First referee}

Dr. Mazur describes molecular dynamics simulations where a correct
static curvature of DNA with phased A-tracts emerges spontaneously
in conditions where any role of counterions or long range electrostatic
effects can be excluded.

I have several problems with this manuscript:

1) The observed curvature is dependent on the starting model. In fact
the manuscript uses the phrase `stable static curvature' incorrectly
to describe what is probably a trapped metastable state. The observed
curve is neither stable nor static.

2) The choice of DNA sequence seems to be biased toward that which
gives an altered structure in simulations, ad is not that which gives
the most pronounced bend in solution. I would suggest a comparison
of (CAAAATTTTTG)n and (CTTTTAAAAG)n.

3) The result is not consistent with solution results. See for example:

Prodin, F., Cocchione, S., Savino, M., \& Tuffillaro, A. ``Different
Interactions of Spermine With a Curved and a Normal DNA Duplex -
(Ca(4)T(4)G)(N) and (Ct(4)a(4)G)(N) - Gel -Electrophoresis and
Circular-Dichroism Studies'' (1992) Biochemistry International 27,
291-901.

Brukner, l, Sucis, S., Dlakic, M., Savic, A., \& Pongor, S.
``Physiological concentrations of magnesium ions induces a strong
macroscopic curvature in GGGCCC - containing DNA'' (1994)
J. Mol. Biol. 236, 26-32.

Diekmann, S., \& Wang, J. C. ``On the sequence determinants and
flexibility of the kinetoplast DNA fragment with abnormal gel
electrophoretic mobilities'' (1985) J. Mol. Biol. 186, 1-11.

Llaudnon, C. H., \& Griffith, J. D. ``Cationic metals promote
sequence-directed DNA bending'' (1987) Biochemistry 26, 3759-3762.

4) The result is not consistent with other simulations. See for
example:

Feig, M., \& Pettitt, B. M. ``Sodium and Chlorine ions as part of the DNA
solvation shell'' (1999) Biophys. J. 77, 1769-81.

5) The results should be given by objective statistical descriptions
rather than a series of spot examples, as in ``sometimes reveals two
independent bends''.

\subsubsection {Second referee}

This manuscript describes the modeling of a 25-residue DNA duplex
using molecular dynamics simulations. The DNA sequence in question
contains 3 A/T tracts arranged in-phase with the helix screw and thus
is expected to manifest intrinsic bending. Unlike previous MD studies
of intrinsically bent DNA sequences, these calculations omit explicit
consideration of the role of counterions. Because recent crystallographic
studies of A-tract-like DNA sequence have attributed intrinsic bending to
the localization of counterions in the minor groove, the present finding
that intrinsic bending occurs in the absence of explicit counterions
is important for understanding the underlying basis of A-tract-dependent
bending.

Overall, the MD procedure appears sound and the calculations were
carried out with obvious care and attention to detail. There are two
specific issues raised by this study that should be addressed in
revision, however.

1. Although the sequence chosen for this study was based on a canonical,
intrinsically-bent motif consisting of three A tracts, it is unclear to
what extent intrinsic bending has been experimentally  shown for this
particular sequence. There are known sequence-context effects that
modulate A-tract-dependent bending and thus the author should refer
the reader to data in the literature or show experimentally that
intrinsic bending of the expected magnitude occurs for this
particular sequence. Moreover, one A tract is out-of-phase with respect
to the others and it is therefore not clear how this contributes to
the overall bend. The author is understandably concerned about end effect
with short sequences; this problem can be ameliorated by examining DNA
fragments that constrain multiple copies of the chosen motif
or by extending the ends of the motif with mixed-sequence DNA.

2. Notwithstanding the authors remark bout separating the cause and
the effects with respect to intrinsic bending some comments about
the underlying mechanism of bending seem appropriate. It would
be particularly useful to know whether average values of any specific
conformational variables are unusual or whether strongly bent states
are consistent with narrowing of the minor groove within A-tracts,
for example.


\begin{thebibliography}{10}

\bibitem{Marini:82}
Marini, J.~C., Levene, S.~D., Crothers, D.~M. \& Englund, P.~T., {\em Proc.
  Natl. Acad. Sci. USA} {\bf 79},  7664--7668  (1982).

\bibitem{Wu:84}
Wu, H.-M. \& Crothers, D.~M., {\em Nature} {\bf 308},  509--513  (1984).

\bibitem{Trifonov:80}
Trifonov, E.~N. \& Sussman, J.~L., {\em Proc. Natl. Acad. Sci. USA} {\bf 77},
  3816--3820  (1980).

\bibitem{Levene:83}
Levene, S.~D. \& Crothers, D.~M., {\em J. Biomol. Struct. Dyn.} {\bf 1},
  429--435  (1983).

\bibitem{Calladine:88}
Calladine, C.~R., Drew, H.~R. \& McCall, M.~J., {\em J. Mol. Biol.} {\bf 201},
  127--137  (1988).

\bibitem{Crothers:99}
Crothers, D.~M. \& Shakked, Z.,  in {\em Oxford Handbook of Nucleic Acid
  Structure}, edited by Neidle, S. (Oxford University Press, New York, 1999),
  pp.\ 455--470.

\bibitem{Zhurkin:79}
Zhurkin, V.~B., Lysov, Y.~P. \& Ivanov, V.~I., {\em Nucl. Acids Res.} {\bf 6},
  1081--1096  (1979).

\bibitem{Sanghani:96}
Sanghani, S.~R., Zakrzewska, K., Harvey, S.~C. \& Lavery, R., {\em Nucl. Acids
  Res.} {\bf 24},  1632--1637  (1996).

\bibitem{Kitzing:87}
von Kitzing, E. \& Diekmann, S., {\em Eur. Biophys. J.} {\bf 14},  13--26
  (1987).

\bibitem{Chuprina:88}
Chuprina, V.~P. \& Abagyan, R.~A., {\em J. Biomol. Struct. Dyn.} {\bf 1},
  121--138  (1988).

\bibitem{Zhurkin:91}
Zhurkin, V.~B., Ulyanov, N.~B., Gorin, A.~A. \& Jernigan, R.~L., {\em Proc.
  Natl. Acad. Sci. USA} {\bf 88},  7046--7050  (1991).

\bibitem{Mirzabekov:79}
Mirzabekov, A.~D. \& Rich, A., {\em Proc. Natl. Acad. Sci. USA} {\bf 76},
  1118--1121  (1979).

\bibitem{Levene:86}
Levene, S.~D., Wu, H.-M. \& Crothers, D.~M., {\em Biochemistry} {\bf 25},
  3988--3995  (1986).

\bibitem{Strauss:94}
Strauss, J.~K. \& Maher, L.~J., III, {\em Science} {\bf 266},  1829--1834
  (1994).

\bibitem{Travers:95}
Travers, A., {\em Nature Struct. Biol.} {\bf 2},  264--265  (1995).

\bibitem{McFail-Isom:99}
McFail-Isom, L., Sines, C.~C. \& Williams, L.~D., {\em Curr. Opin. Struct.
  Biol.} {\bf 9},  298--304  (1999).

\bibitem{Chiu:99}
Chiu, T.~K., Zaczor-Grzeskowiak, M. \& Dickerson, R.~E., {\em J. Mol. Biol.}
  {\bf 292},  589--608  (1999).

\bibitem{Young:98}
Young, M.~A. \& Beveridge, D.~L., {\em J. Mol. Biol.} {\bf 281},  675--687
  (1998).

\bibitem{Mzjmb:99}
Mazur, A.~K., {\em J. Mol. Biol.} {\bf 290},  373--377  (1999).

\bibitem{Dickerson:89}
Dickerson, R.~E. {\it et~al.}, {\em J. Mol. Biol.} {\bf 205},  787--791
  (1989).

\bibitem{Mzjbsd:89}
Mazur, A.~K. \& Abagyan, R.~A., {\em J. Biomol. Struct. Dyn.} {\bf 6},
  815--832  (1989).

\bibitem{Mzjcc:97}
Mazur, A.~K., {\em J. Comput. Chem.} {\bf 18},  1354--1364  (1997).

\bibitem{Mzjchp:99}
Mazur, A.~K., {\em J. Chem. Phys.} {\bf 111},  1407--1414  (1999).

\bibitem{Mzjacs:98}
Mazur, A.~K., {\em J. Am. Chem. Soc.} {\bf 120},  10928--10937  (1998).

\bibitem{Mzlanl:99}
Mazur, A.~K., {\em Preprint} {\bf {\rm http: // xxx.lanl.gov/abs/
  physics/9907028}},    (1999).

\bibitem{AMBER94:}
Cornell, W.~D. {\it et~al.}, {\em J. Am. Chem. Soc.} {\bf 117},  5179--5197
  (1995).

\bibitem{Cheatham:99}
Cheatham, T.~E., III, Cieplak, P. \& Kollman, P.~A., {\em J. Biomol. Struct.
  Dyn.} {\bf 16},  845--862  (1999).

\bibitem{TIP3P:}
Jorgensen, W.~L., {\em J. Am. Chem. Soc.} {\bf 103},  335--340  (1981).

\bibitem{Arnott:72}
Arnott, S. \& Hukins, D. W.~L., {\em Biochem. Biophys. Res. Communs.} {\bf 47},
   1504--1509  (1972).

\bibitem{Curves:}
Lavery, R. \& Sklenar, H., {\em J. Biomol. Struct. Dyn.} {\bf 6},  63--91
  (1988).

\end{thebibliography}
\end{document}